\documentclass[preprint]{aastex}
\shortauthors{THORSTENSEN, L\'EPINE, \& SHARA}
\shorttitle{Parallaxes of Cataclysmics}
\begin{document}
\title{Parallax and Distance Estimates for Twelve Cataclysmic Variable Stars
\footnote{Based on observations obtained at the MDM Observatory, operated by
Dartmouth College, Columbia University, Ohio State University,
the University of Michigan, and Ohio University.}
}

\author{John R. Thorstensen}
\affil{Department of Physics and Astronomy\\
6127 Wilder Laboratory, Dartmouth College\\
Hanover, NH 03755-3528;\\
john.thorstensen@dartmouth.edu}

\author{S\'ebastien L\'epine and Michael Shara}
\affil{Department of Astrophysics, Division of Physical Science\\
American Museum of Natural History\\
Central Park West at 79th Street\\
New York, NY 10024}

\begin{abstract}
We report parallax and distance estimates for twelve more 
cataclysmic binaries and related objects observed with the
2.4m Hiltner telescope at MDM Observatory.  The final
parallax accuracy is typically $\sim 1$ mas.  
%For only one of the twelve objects, IR Gem, do
%we fail to detect a significant parallax. 
Notable results include distances for 
V396 Hya (CE 315), a helium double degenerate with a relatively long 
orbital period, and for MQ Dra (SDSSJ155331+551615), a magnetic system 
with a very low accretion rate.  We find that the Z Cam star KT Persei 
is physically paired with a K main-sequence star lying 15 arcsec away.
Several of the targets have distance estimates in the literature that are
based on the white dwarf's effective temperature and flux; our measurements
broadly corroborate these estimates, but tend to put the stars a bit 
closer, indicating that the white dwarfs may have rather larger masses
than assumed.
As a side note, we briefly describe radial velocity spectroscopy that refines
the orbital period of V396 Hya to $65.07 \pm 0.08$ min.
\end{abstract}
% V884 Her is just about ready, too.

\section*{}
\begin{quote}
\textit{``\ldots Obs of Sirius must be taken as far apart as possible, mustn't they,---
at least six months of what the World no doubt sees as Idleness, whilst the 
Planet, in its good time, cranketh about, from one side of its Orbit to 
the other, the Base Line creeping ever longer, the longer the 
better$\ldots$how is any of that my fault?''} -- Neville Maskelyne, as
imagined by Thomas Pynchon in {\it Mason \& Dixon}. \citep{Pynchon}
\end{quote}

\section{Introduction}

Cataclysmic variables (CVs) are interacting binary star systems 
in which a white dwarf accretes matter from a close companion, which usually resembles
a main-sequence star.  \citet{warn} has written an excellent 
monograph on CVs.  

This is the second in a series of papers presenting parallax and 
distance estimates for selected CVs and related objects.  The introduction
to the first paper (\citealt{thorstensen03}; hereafter Paper I) outlines 
some of the
indirect methods used to estimate CV distances, and the situation up until that
point.  Since then, more parallaxes have been measured using
the Fine Guidance Sensor (FGS) on the Hubble Space Telescope (HST)
\citep{harrison04,exhya,v1223sgr,amcvns}.  
% insert reference to AM CVn stars
While the FGS has the best 
astrometric precision of any instrument currently available 
(save for VLBI at radio frequencies; \citealt{reid07}),
the number of parallaxes that can be measured with FGS is sharply limited.  
Fortunately, parallaxes accurate to $\sim 1$ milliarcsec (mas) 
can be obtained from the ground (\citealt{monet92}; Paper I).  

The parallax program has been expanded to include $\sim 60$ stars
discovered in the L\'epine-Shara Proper Motion survey \citep{lspm},
headquartered at the American Museum of Natural History (AMNH).
AMNH and Dartmouth both have access to the MDM telescopes, so we
are coordinating our efforts and obtaining more frequent observations.
Results on these objects will be published elsewhere.

In Section 2, we briefly summarize our procedures, with emphasis on 
refinements implemented since Paper I.  Section 3 details the measurements
of the individual stars and their distances.  In Section 4, we
present a summary and discussion.

\section{Procedures and Targets}

The observation and reduction protocols used for this study
are nearly identical to those described in Paper I.
We summarize that lengthy discussion here for the convenience of the reader.

All the images are from the 2.4 m Hiltner telescope at MDM Observatory.
A 2048$^2$ SITe CCD detector is used, with the readout cropped to 
1760$^2$ because of vignetting by the 50 mm filters used.  The
24 $\mu$m pixels project to $0''.275$ on the sky.  An $I$-band filter
is used for all the parallax observations; this minimizes
differential color refraction (DCR) effects \citep{monet92}.
%Some brighter objects have recently been added to the program using a 
%$\lambda 7000$ narrow-band interference filter kindly loaned by 
%R. Fesen, but we do not report any results from this filter here.
Exposure times are for the most part fairly short, typically 90 s, in order
to avoid saturating the program object or the main reference stars,
but this varies from field to field.  We center the image
on the object to within a few arcseconds in order to keep a consistent
set of reference stars in the field of view.
In a typical visit to an object, 6 to 10 exposures are obtained to 
average out the astrometric noise and to permit rejection of images
that show especially large astrometric scatter.  
Median images of the twilight sky serve for flat field correction.  
When conditions are clear, we obtain
one or two images of the program field in the $V$ filter,
and also observe standard star fields selected from \citet{landolt92} 
to derive $I$ magnitudes and $V-I$ colors for the program and 
reference stars.

Reduction and analysis proceed almost exactly as described in Paper I, but
with greater automation of the routine steps.  The reduction and
analysis routines make heavy use of the scripting capabilities
of the PyRAF\footnote{PyRAF is a Python-language wrapper for
the IRAF package, written at Space Telescope Science Institute.
It is available at {\tt http://www.stsci.edu/resources/software\_hardware/pyraf}.
IRAF, the Image Reduction and Analysis Facility, is distributed by the
National Optical Astronomy Observatory, which is operated under contract
to the National Science Foundation.} package.

Since Paper I we have measured the differential color
refraction more precisely than before.  
In that paper we used a value of 7 mas per unit $V-I$ per
unit $\tan Z$ for the DCR coefficient, but a re-measure gives
a value of 5 in the same units.  This also agrees better with the 
value we had calculated using stellar energy distributions and 
the atmospheric dispersion curve (see Paper I).  Adopting the new 
value makes little difference in our results, since we take 
pains to observe within $\pm 2$ hr of the meridian -- more usually 
$\pm 1$ hr -- where the effect of DCR should be relatively small
and consistent.

The astrometric reduction yields the relative parallax 
$\pi_{\rm rel}$ and the relative proper motions $[\mu_X,\mu_Y]$ --
all these quantities are relative to the reference stars chosen.
We make no attempt to correct the proper motions to an inertial
frame, but we do estimate the mean parallax of the reference stars
using their magnitudes and colors (see Paper I), resulting in a
correction to $\pi_{\rm rel}$; typically the estimated absolute
parallax $\pi_{\rm abs}$ is larger than $\pi_{\rm rel}$ by 
$\sim 1$ mas.

In the last step of our analysis we use a Bayesian procedure that in 
principle generates an overall best estimate of the distance, taking 
into account all the available information.  The procedure includes 
(a) a Lutz-Kelker correction, which accounts for the bias 
introduced by the rapidly increasing volume of space represented
as the parallax becomes small; (b) a distance weighting based on the 
observed proper motion {\it and} an {\it assumed} velocity 
distribution for the CV population; (c) any prior distance 
estimates based on independent indicators.  These include
estimates based on the spectral type and apparent magnitude of the 
secondary, the solid angle
subtended by the white dwarf as deduced from UV spectrophotometry,
and other methods.  When the relative precision of the parallax is good,
the prior information makes little difference, but it becomes more
important as the parallax precision degrades.  Paper I discusses
the Bayesian procedure at length.

Although we have not significantly changed our analysis procedures, 
we have noted some effects with more experience.  

We are more aware of how strongly the photon
statistics affect the centroiding accuracy.  The brightest stars in the 
field almost always show very little scatter -- less than 5 mas
vector error in a single exposure in the best cases -- while
faint stars show very large scatter.   We are now
more careful to select longer exposure times for the fainter
objects (which, unfortunately, overexposes some bright
potential reference stars). 

We have also found that the correlation between the seeing,
as measured by the full-width half maximum (FWHM) of the point-spread 
function (PSF), correlates only approximately with the 
astrometric quality of the image, as 
judged from the scatter in the the reference star positions.  Useful
astrometry can come from
mediocre seeing, and images taken in good seeing can sometimes show
surprisingly large astrometric scatter.  When the seeing is 
bad enough, the expected correlation does appear -- images with
very poor seeing (FWHM $> 2.5$ arcsec) are essentially useless for
parallax.

Finally, we have found that typically at least two seasons of 
data are needed in order for us to feel confident that the 
astrometric solution has `settled',
because the proper motion in each axis needs to be measured
independently of the parallax.  The statement attributed to 
Maskelyne in this paper's epigraph is evidently too
optimistic.

We selected targets using informal criteria, which included the
following:  (a) We found a distance estimate in the literature,
which suggested that the parallax would be detectable, or that it was 
astrophysically interesting to establish a lower limit on the distance. 
(b) The proper motion was relatively high.  (c) The apparent
magnitude was relatively bright for the subclass.  (d) A parallax was
requested by a colleague.   

Table 1 lists the stars observed, ordered by constellation and variable
star name, and the epochs at which we observed them.   
We discuss each star in turn.

\section{Results}

Table 2 lists positions, magnitudes, relative proper motions, and
relative parallaxes for {\it all} the measured stars in all the fields.   
The proper motions are relative to the mean of all the stars in the field; 
they have not been reduced to a non-rotating reference frame. 
Table 3 lists the parallaxes and distance estimates for the
program stars.  The different distance estimates in Table 3
take into account different amounts of Bayesian prior information.

We now discuss the individual stars in turn, ordered as they are
in the GCVS (alphabetically by full constellation name; within
constellations, by variable star designation; \citealt{gcvs}).

\subsection{DW Cnc}

DW Cnc is a remarkable CV, unique in that the radial velocities
of its very strong emission lines are modulated on two distinct, 
incommensurate periods, 86.1015(3) and 69.9133(10) minutes, each of
these periods having similar velocity amplitude.  The light curve
is modulated at the beat frequency of the two velocity periods.  
The three periods are most readily explained as the orbital period $P_{\rm orb}$, 
the rotation period of a magnetized white dwarf, and the orbital 
sideband \citep{pattersondw, rodriguezdw}.  

Although the parallax is determined with good precision, the relative
error $\sigma_\pi/\pi$ is substantial, leading to a sizeable Lutz-Kelker
correction.  Because the system is unusual, we unweighted the
distance prior in the Bayesian 
analysis by assuming a large uncertainty in luminosity.  We
find a proper motion $[\mu_X,\mu_Y] = [-26, +5]$ mas yr$^{-1}$,
consistent with the USNO B value of $[-22 \pm 5, +0 \pm 2]$ \citep{usnob}. 
Our final distance estimate, $257 (+79, -52)$ pc, puts DW Cnc near 
the peak of prior distance probability based on the proper motion
alone, which in turn is based on the observation that most
CVs have modest space velocities (Paper I).

\citet{pattersondw} reported extensive photometry giving
$V = 14.8$ for DW Cnc, which corresponds to $M_V = 7.8$ at 250 pc.  
This is within the range of values predicted by the 
\citet{warn87} relation for an outbursting
dwarf nova at $P_{\rm orb} = 86$ min; literal agreement
would require an inclination $i \sim 83$ degrees.
An inclination this large is ruled out by the lack of eclipses.
The \citet{warn87} relation was derived for dwarf novae at maximum light, 
and DW Cnc is not a dwarf nova;  nonetheless, DW Cnc is clearly much 
more luminous than short-period dwarf novae at minimum light.
It is evidently a rare example of a novalike variable with
$P_{\rm orb} < 2$ hr.

\subsection{HT Cas}

HT Cas is a well-studied, deeply-eclipsing dwarf nova of the SU UMa
subclass, with $P_{\rm orb} =  106$ min.  \citet{feline} present 
recent high-speed photometry and summarize some of the earlier work on
this object.  Because HT Cas eclipses, there is a long history of 
detailed time-resolved studies aimed at mapping the disk emission and
constraining the white dwarf radius, which has led to
several estimates of the distance.  \citet{marshhtcas} detected the
secondary star, classified it as M5.4 $\pm$ 0.3, and measured
the secondary's contribution to total light at eclipse.  
\cite{horne91} derived dynamical constraints using high-speed
photometry of the eclipses; from these, and the secondary's
contribution, they estimated a distance of 135 pc.  \citet{vrielmannht}
give a critical discussion of previous distance estimates, and
using a different mapping technique, derive a distance of 
207 $\pm$ 10 pc.

The parallax measurement decisively favors the shorter distance;
the \citet{vrielmannht} distance is essentially excluded.  While the
$\sim 130$ pc distance favored by \citet{marshhtcas} 
is a little longer than our astrometric distance, it
lies well within the uncertainties.  Our final Bayesian distance
estimate includes prior distance information; for this we
adopted a rough average value of previous estimates and 
assumed an uncertainty generous enough to includes both 
the longer and shorter distances, namely
$m - M = 6.1 \pm 1$, or $166 (+97,-61)$ pc.  The parallax
is accurate enough that this assumption increases the 
distance estimate by only 8 pc.

The proper motion found here,
$[\mu_X,\mu_Y] = [+27, -14]$ mas yr$^{-1}$, agrees well with
that listed in the Lick NPM2 catalog, $[+28,-10]$ mas yr$^{-1}$ 
\citep{licknpm2}. 
Assuming that the distance is at the far end of our
1$\sigma$ error bar, the transverse velocity referred to the 
LSR is only $\sim 6$ km s$^{-1}$.  Due to the faintness
of the secondary, we were unable to find a precise systemic 
radial velocity $\gamma$ in the literature, but 
emission-line studies 
suggest that $\gamma$ is some tens of 
km s$^{-1}$ or less \citep{marshhtcas,voracious}.  The 
kinematics of HT Cas are clearly those of the disk.  

\subsection{MQ Dra}

MQ Dra was discovered in the Sloan Digital Sky Survey
\citep{sdsscv2}, in which it was called 
SDSS J155331.12+551614.5.  The spectrum shows a
white dwarf with a $\sim 60$ MG magnetic field, 
and only weak emission features, indicating a very 
low accretion rate \citep{szkodylarp}.  It is 
therefore a fine example of a low accretion rate polar, or 
LARP (see, e.g., \citealt{schmidtlarp05}).  
It has $P_{\rm orb}$ = 4.39 h, and the secondary star's
spectral type is $\sim$M4.5 \citep{harrisonmq}.  
The secondary's contribution to the light leads
to a distance estimate of 130 to 180 pc 
\citep{schmidtlarp05, szkodymq}.
%the distance is 
%M5V secondary to have a luminosity similar to field
%main sequence stars of the same spectral type, 
%\citet{szkodylarp} estimated a distance modulus
%$m - M = 5.1$, corresponding to $d = 104$ pc.  

Although the field of MQ Dra is somewhat sparse, 
the parallax is measured with a formal  
uncertainty of only 0.7 mas; the scatter of 
apparent field star parallaxes suggests that this 
is a realistic uncertainty, but we adopt 
1.0 mas to be conservative and to allow for the 
unmodeled uncertainty in the 
correction from $\pi_{\rm rel}$ to $\pi_{\rm abs}$.
Our measured relative proper motion 
$[\mu_X,\mu_Y] = [-31, +6]$ mas yr$^{-1}$, 
agrees well the proper motion listed in the 
USNO B, which is $[-28 \pm 5, -8 \pm 3]$
in the same units \citep{usnob}.   The parallax is accurate enough that
the various corrections do not strongly affect the 
distance estimate.  For the final Bayesian estimate,
we include a secondary-based distance modulus
corresponding to $d = 150 (+60, -40)$ pc.  Combining 
this with the parallax and proper motion yields
$d = 162 (+27, -21)$ pc for the final estimate.

\subsection{IR Gem}

IR Gem is an SU UMa-type dwarf nova with $P_{\rm orb} = 98.5$ min
\citep{feinswogirgem,irgemphot}.
The parallax is so small that it is barely detected,
despite the large and favorable set of comparison stars available
at IR Gem's Galactic latitude.  The proper motion is sizeable;
we measure $[\mu_X,\mu_Y] = [+53.9,-22.3]$ mas yr$^{-1}$,
in good agreement with the UCAC2 catalog \citep{ucac2}, which gives $[+51.9 \pm 3.4, 
-28.8 \pm 3.4]$ mas yr$^{-1}$.  Because of the small parallax, the
Lutz-Kelker correction depends critically on the uncertainty 
adopted, which we estimate conservatively using the scatter in 
the fitted parallaxes of other stars in the field (see Paper I).  

In the case of IR Gem, $\pi_{\rm abs}/\sigma_{\pi} < 4$, so there 
is no local maximum in the Lutz-Kelker parallax probability function;
the parallax alone does not formally give a distance.
With such weak parallax evidence, the
Bayesian priors steer the distance estimate.
For a distance prior, we use the \citet{warn87} relations.
The General Catalog of Variable Stars (GCVS; \citealt{gcvs}) lists
$V = 10.7$ at maximum light.  Since superoutbursts are 
$\sim 1$ magnitude brighter than the normal outbursts for 
which the \citet{warn87} relation is calibrated, we adopt
$V = 11.7$ in computing the distance prior.  Applying the
relation requires an inclination estimate; \citet{feinswogirgem}
note that the radial velocity amplitude is low, which suggests
an orbital inclination not too far from face-on.  At
low inclinations, the correction is fortunately nearly constant;
we arbitrarily adopt $i = 30$ degrees, leading to a 
prediction of $M_V({\rm max}) = 4.6$ for $P_{\rm orb} = 98$ min.
The resulting distance is $\sim 250$ pc.  
We adopt an uncertainty of $\pm 1$ mag in the distance modulus
for the Bayesian prior.

The Bayesian results are as follows.
The large proper motion, interpreted using the space velocity 
probability density assumed in Paper I, eases the Lutz-Kelker problem
and results in a 50-th percentile distance near 410 pc.
The prior distance estimate decreases this to $358 (+ 104, -86)$ 
pc.  At 358 pc, the transverse velocity corresponds to
98 km s$^{-1}$ referred to the LSR; including a systemic
radial velocity $\gamma = +60$ km s$^{-1}$ (estimated from 
\citealt{feinswogirgem})
gives a space velocity of $\sim 115$ km s$^{-1}$.  This puts 
IR Gem out on the high-velocity tail of the assumed
CV velocity distribution.

\subsection{V396 Hya = CE 315}

V396 Hya is an AM CVn-type system, in which the secondary
is evidently an evolved degenerate star with no hydrogen
remaining.  It was discovered
as a high proper motion star in the Calan-ESO survey \citep{calan}, and
designated CE 315 \citep{ce315}.  Its orbital
period, 65 min, is the longest of all
known AM CVn systems.  Systems with degenerate
secondaries tend to evolve toward longer orbital periods,
so V396 Hya is likely to be among the most highly evolved
AM CVn stars known.  \citet{bildsten} found that, for these
relatively long-period AM CVn stars, the luminosity
arises largely from the accreting white dwarf, and 
predicted a system luminosity.
They used a preliminary version of our
parallax measurement to show that the 
luminosity of V396 Hya agrees well with theory.
At $V = 17.6$, V396 Hya is just a little too faint 
for the HST FGS.

In addition to our parallax determination, 
we have some spectra of V396 Hya taken
with the Hiltner telescope and modular 
spectrograph; the instrumentation, procedures, and 
reductions were essentially as described by \citet{longp}.
We took 35 exposures, most of them 5 min, on the 
nights of 2001 May 14, 15, 16, and 17 UT.  The average
spectrum appeared very similar to that shown by
\citet{ce315}.  We measured radial velocities of the wings
of the HeI $\lambda$5876 emission line using a double-gaussian 
convolution algorithm \citep{sy80, shafter83}, and 
searched for periods using a `residual-gram' algorithm
\citep{tpst}.  The \citet{ce315} period was clearly
detected as the strongest among several aliases 
(which are inevitable because the observations are
clustered modulo 24 hours).  By fitting the velocities
from all the nights, we refine $P_{\rm orb}$ to $65.07 \pm 0.08$ min.
The emission line profiles in our average spectrum
are mostly triple-peaked, as described by \citet{ce315}.
In an effort to constrain the systemic radial
velocity $\gamma$, we fit the tops of the central
peaks with Gaussians using the interactive {\tt splot} command 
in IRAF.  The different lines gave velocities ranging from 
+30 to +70 km s$^{-1}$ (heliocentric); from a weighted average, we
estimate $v = +45 \pm 20$ km s$^{-1}$ for the line core.
It is entirely possible that this does not accurately reflect 
the systemic velocity.

The parallax is determined to good relative accuracy and 
dominates the Bayesian estimate of distance.  We find
a proper motion of $[-277,-52]$ mas yr$^{-1}$, 
significantly smaller than the Calan-ESO value, but 
in reasonable agreement with the USNO B, which 
lists $[-264,-30]$ mas yr$^{-1}$ \citep{usnob}.  This large proper
motion more than compensates for the Lutz-Kelker
effect in the Bayesian calculation, so the 
50th-percentile distance of 92 pc is essentially equal to 
$1/\pi_{\rm abs}$.  The final estimate is 
slightly larger than the 77 pc preliminary estimate 
used by \citet{bildsten}, but their conclusions should
be essentially unchanged.  Some insight into the
kinematics of this system can be had by computing 
the tangential velocity, referring it to the LSR, 
and rotating into Galactic coordinates; for a
distance of 90 pc, this yields 
$[U,V,W] = [+79,-68,0]$ km s$^{-1}$ --
oddly enough, nearly
parallel to the plane of the Galaxy.  If we adopt
$\gamma = +45$ km s$^{-1}$, this becomes
$[+57,-95,+28]$ km s$^{-1}$.  In any case, these are 
not the kinematics of the thin disk (though the system 
apparently does not
stray far from the plane).  It is interesting to note that
another AM CVn system, GP Com, evidently also has
a high space velocity \citep{thorstensen03}.   It may 
be significant that both V396 Hya and GP Com, two
of only $\sim$18 known AM CVn stars \citep{roelofsAM}, 
also belong in the small handful of CVs that were first 
discovered because of their proper motions.  This suggests that 
AM CVn systems are fairly common, or that many are 
members of an old population, or both.  On the other
hand, \citet{amcvns} suggest
that the space density is not particularly large, based on 
the HST parallaxes of a sample of AM CVn stars.

\subsection{KT Per}

KT Per is a Z Cam-type dwarf nova with $P_{\rm orb}$ 
near 3.90 hr \citep{ratering, thorktper}.  We chose it
for parallax determination because of its substantial
proper motion in the UCAC2, $[\mu_X,\mu_Y] = [+63,-5]$ mas
yr$^{-1}$ \citep{ucac2}.  Also, \citet{thorktper} estimated a distance
of $245 \pm 100$ pc based on the M3 secondary star, 
which is close enough for parallax to provide a useful check. 

Our astrometry shows that the $V = 15.19$ field
star located 15.4 arcsec west and 2.3 arcsec north of
KT Per is evidently a physical companion, sharing with KT Per
a nearly identical large proper motion and parallax. 
We obtained spectra of this star with the Hiltner 2.4m 
telescope and calibrated them using observations of flux
standards.  We have a library of spectra of K dwarfs classified by
\citet{keenan89}, taken with the same instrument.  Using these
for comparison, we estimate the spectral type of the companion
to be K$4.5 \pm 1$ subclass.  \citet{pickles} tabulates 
mean colors as a function of spectral type.   We measure
$V-I = +1.59$ for this star, which, by his table, would indicate a 
K7 to M0 star, significantly later than observed.  However, 
\citet{ladous} estimated  $E(B-V) = +0.2$ 
for KT Per, using the strength of the 2200 \AA\ extinction
feature in archival IUE spectra.  Combining this with 
$E(V-I)/E(B-V) = 1.35$ \citep{hereddening}
yields $(V-I)_0 = 1.32$, which is nicely consistent with the
spectral type.  As a check, we note the reddening maps of
\citet{schlegeldust} give $E(B-V) = 0.32$ for the 
total Galactic extinction in this direction 
($l = 130^{\circ}.2, b = -11^{\circ}.3$); it is not
implausible that KT Per is extinguished by a substantial 
fraction of this total.  Taking Pickles' mean value
of $M_V = 7.64$ for this star, and taking $A_V = 
3.2 E(B-V)$, yields a spectroscopic distance modulus
$(m - M)_0 = 6.9$, corresponding to 240 pc, in close
agreement with the somewhat cruder estimate from the 
secondary star.  Using cross-correlation methods 
described in \citet{longp},  we find the companion's
heliocentric radial velocity to be 
$-2 \pm 10$ km s$^{-1}$. 

As table \ref{tab:results} shows, the parallax is small, but
clearly detected, and the distance derived from the 
parallax, while broadly consistent with the photometric
estimate, is a bit shorter.  The compromise Bayesian distance
estimate is $180 (+36, -28)$ pc; at 180 pc, 
the space velocity with respect to the LSR is 38 km s$^{-1}$,
not too large for a disk population object.

\subsection{VV Pup}

VV Pup was among the first CVs to be recognized as an AM Herculis star, or polar
\citep{tapia77, bond77}.  
%\citet{mason07} present time-resolved spectroscopy in an 
%extreme low state showing the cyclotron features of the white dwarf.  
\citet{araujo-betancor05} detect the white dwarf in ultraviolet spectra, and
derive distances from model-atmosphere fits.  They fit both
one- and two-component models, giving somewhat different results; the
full range of their estimate can be roughly summarized as 
$169 \pm 35$ pc, where the error bar is largely dominated by the
unknown mass (hence radius) of the white dwarf.  
\citet{howell06} measured radial velocities of the secondary
star in the infrared during a low state, and infer a distance
of $\sim 120$ pc from its K-band magnitude; arbitrarily assigning
this an uncertainty of $\pm 50$ pc and combining it with the 
white-dwarf based distance gives a weighted average of 
$153 \pm 29$ pc for the prior estimate.  The astrometric
distance found here is somewhat shorter than this, namely
$114 (+18, -14)$ pc; combining these leads to a final best guess
distance of $124 (+17, -14)$ pc.   At this distance the 
relative proper motion corresponds to a tangential
velocity of 43 km s$^{-1}$, referred to the LSR.  It is therefore a little 
surprising that \citet{howell06} find a systemic velocity
$\gamma = -130 \pm 18$ km s$^{-1}$ from their observations of the 
secondary star, as that would imply a large space velocity
directed mostly toward us.  Although previous radial-velocity studies
(e.g. \citealt{sy80vv,cowley82,diaz94}) were based on observations
of emission lines in the high state, and hence may not have
measured $\gamma$ accurately, they do not give any indication
of such a large approach velocity.  \citet{howell06} 
point out that their velocities are based on an effective wavelength
they assume for the blended 2.2 $\mu$m sodium doublet;
an inappropriate choice for this would
throw their $\gamma$ velocity off.  Without a reliable $\gamma$, 
we cannot find a full space velocity, but the transverse components
suggest that VV Pup has the kinematics of the disk population.

\subsection{MR Ser}

\citet{liebert82} called attention to PG 1550+191, later named
MR Ser, as the only AM Her star discovered in the Palomar-Green
ultraviolet survey.  \citet{mukaicharles86} detected TiO 
absorption bands and estimated the spectral type of the 
secondary as M5-M6. Using the assumptions that the secondary 
fills its Roche lobe and obeys a zero-age main sequence 
mass-radius relation, they estimated $d = 142$ pc from the
secondary; \citet{schwope93}, using similar methods, estimated
$d = 139 \pm 13$ pc independently.  As with VV Pup, 
\citet{araujo-betancor05} detected the white dwarf in HST
ultraviolet spectra, and again fitted one- and two-component
atmospheres, yielding respectively $180 (+23, -30)$ 
and $160 (+18, -26)$ pc.  Combining all these, we adopt
$155 \pm 25$ pc as the prior distance estimate.  Once again,
the astrometric distance -- $113 (+14, -12)$ pc -- is 
somewhat shorter than the prior estimate.  This suggests that 
the mass of the white dwarf is fairly high, resulting in 
a smaller radius and hence that the solid angle inferred
from the atmosphere fits corresponds to a shorter distance.
The secondary star's surface brightness may also have been 
overestimated.  In any case, the prior estimate increases 
the best-guess distance to $126 (+14, -12)$ pc from its 
purely astrometric value.  

The transverse velocity at the nominal distance is 
44 km s$^{-1}$ referred to the LSR; \citet{schwope93} find
a mean velocity $\gamma = 7 \pm 18$ km s$^{-1}$ for the 
sodium infrared doublet, which tracks the secondary's photosphere and
hence should give a trustworthy reading of the systemic
velocity.  Including this gives an LSR space velocity of 
49 km s$^{-1}$. 

\subsection{SW UMa}

SW UMa is an SU UMa-type dwarf nova with $P_{\rm orb} =  
81.8$ min \citep{shafter86}.  
\citet{gaensicke05} fit white dwarf model atmospheres to 
ultraviolet spectra to find $d = 159 \pm 22$ pc, after allowing
for variation of the white dwarf's assumed mass in the range
0.35 - 0.9 M$_{\odot}$.  We began taking astrometric images
of this target in 1999 January, and then dropped it from the program for
a time before resuming in 2004.
The relative (not absolute!) proper motion is therefore determined 
very precisely, with formal errors near 0.2 mas yr$^{-1}$ in each coordinate.
The parallax is 
detected well and gives a distance in good agreement with the 
prior distance estimate, without substantially improving
on its formal precision.  The proper motion is fairly modest
and implies an LSR transverse velocity of 25 km s$^{-1}$ at the 
nominal distance.  Combining this with 
$\gamma = -10$ km s$^{-1}$ (estimated from Fig.~6a of 
\citealt{shafter86}) increases this only slightly to 
27 km s$^{-1}$, comfortably consistent with disk-star
kinematics.

\subsection{AR UMa}

This AM Her-type system has an 
unusually strong white-dwarf magnetic field of $\sim 230$ MG
\citep{schmidt96, hoard04}.
\citet{remillard94} found $P_{\rm orb} = 1.93$ h and
detected features from the secondary star, which they classified
as M6.  Based on this, they estimated $d = 88 \pm 18$ pc.
The present measurement agrees closely with this,
yielding $d = 86 (+10, -8)$ pc from the parallax and kinematic
information alone; including the \citet{remillard94} estimate 
does not change this significantly.  AR UMa is evidently one of the 
nearest magnetic CVs.  At 86 pc the transverse velocity referred
to the LSR is 16 km s$^{-1}$.

\subsection{BZ UMa}

BZ UMa is an SU UMa-type dwarf nova.  \citet{ringwald94} found 
$P_{\rm orb}$ = 98 min and detected an M5.5 $\pm$ 0.5 secondary, 
from which they estimated $d = 110 (+44, -51)$ pc; however,
a $JHK$ measurement yielded an alternate constraint,
$d > 140$ pc.   The reference
frame is rather sparse, and the parallax is only just detected; 
the astrometric distance (including proper motion) is 
$260 (+93,-58)$ pc.  Mixing in the 
secondary-based distance modulus, unweighted by a generous 1.4 magnitude
estimated uncertainty, brings the distance estimate down
to $228 (+56, -38)$ pc.  At this distance, the modest proper 
motion corresponds to a tangential velocity of 35 km s$^{-1}$ referred
to the LSR.

\subsection{QS Vir = EC13471-1258}

\citet{odonoghue03} published an extensive study of this system,
which is an eclipsing WD + dMe system with a period of 
3$^{\rm h}$ 37$^{\rm m}$.  While not showing clear evidence of 
mass transfer at present, the system appears consistent with a
hibernating CV, with a secondary just filling its Roche lobe.
The favorable circumstances allowed \citet{odonoghue03} to 
determine the mass and radius of the white dwarf to be $0.78 \pm 0.04$
M$_{\odot}$ and $0.011 \pm 0.001$; a white-dwarf model atmosphere 
yielded a distance of $48 \pm 5$ pc.  The measurement here was
undertaken as an independent check of this measurement.  The 
astrometric distance proves to be $49 (+5, -4)$ pc, in near-perfect
agreement with their result, but unfortunately not significantly
refining the distance estimate.

\section{Conclusion}

We have measured, or at least constrained, the distances to a sample of twelve
CVs.  As with Paper I, we do not find any striking discrepancies with previous
work.   Our most significant results are as follows:  
\begin{enumerate}
\item{The luminosity of DW Cnc is relatively high for its short orbital
period, making it an unusual short-period novalike system.}
\item{The parallax of HT Cas decisively favors the shorter of two possible
distance ranges in the literature.}
\item{V396 Hya is the first long-period AM CVn star with a known distance.}
\item{KT Per forms a wide physical pair with a K dwarf.  This may be useful
as a proxy for estimates of the system's age and metallicity.}
\item{Our distance for QS Vir is in essentially perfect agreement with 
a prior estimate by \citet{odonoghue03}.}
\item{For VV Pup and MR Ser, both of which are magnetic systems, 
our distances are somewhat shorter than those determined from white dwarf 
model atmosphere fits.  The discrepancy is not too serious, but it may 
indicate that the white dwarfs in these systems are more massive than 
had been assumed.}
\end{enumerate}

\acknowledgments
We thank Darragh O'Donoghue, Tom Harrison, Boris G\"ansicke, and 
Joe Patterson for suggesting
targets.  Dartmouth graduate students Chris Peters and Cindy Taylor 
obtained some of the observations.  The staff of MDM Observatory gave their
usual excellent support, and an anonymous referee provided a timely and helpful 
report.  We are always thankful to the Tohono O'odham Nation for
leasing us their mountain, so that we may explore the universe that
surrounds us all.  This research was supported by the 
National Science Foundation through grants AST-0307413 and AST-0708810
at Dartmouth College and AST-0607757 at the American Museum of Natural
History..
\clearpage

% References

% Figures with figure legends (if any)

% Tables (if any)

\begin{deluxetable}{lrrrl}
\tabletypesize{\scriptsize}
\tablewidth{0pt}
\tablecolumns{5}
\tablecaption{Journal of Observations}
\tablehead{
\colhead{Star} & 
\colhead{$N_{\rm ref}$} & 
\colhead{$N_{\rm meas}$}  & 
\colhead{$N_{\rm pix}$} &
\colhead{Epochs} \\ 
}
\startdata

  DW Cnc  &   37  &  80  & 106  & 2004.03(17) , 2004.16(17) , 2004.87(7) , 2005.21(2) , 2005.30(9) , \\
           &       &      &      &  2005.88(12) , 2006.04(15) , 2006.19(8) , 2006.84(7) , 2007.07(4) , \\
           &       &      &      &  2008.05(8) \\
  HT Cas  &   85  & 158  & 130  & 2004.03(15) , 2004.86(11) , 2005.71(15) , 2005.88(11) , 2006.04(12) , \\
           &       &      &      &  2006.64(18) , 2006.84(14) , 2007.07(6) , 2007.65(5) , 2007.73(7) , \\
           &       &      &      &  2007.91(7) , 2008.05(9) \\
  MQ Dra  &   29  &  62  & 124  & 2003.46(17) , 2004.16(15) , 2004.47(27) , 2005.21(12) , 2005.30(4) , \\
           &       &      &      &  2005.48(14) , 2005.70(5) , 2006.20(2) , 2006.38(6) , 2006.44(7) , \\
           &       &      &      &  2007.23(9) , 2007.48(6) \\
  IR Gem  &   48  &  75  & 174  & 2003.08(10) , 2004.03(22) , 2004.16(23) , 2004.87(9) , 2005.08(9) , \\
           &       &      &      &  2005.20(8) , 2005.88(11) , 2006.04(25) , 2006.20(15) , 2006.83(10) , \\
           &       &      &      &  2007.07(8) , 2007.24(7) , 2008.04(9) , 2008.15(8) \\
V396 Hya  &   43  &  82  & 199  & 2003.08(17) , 2003.46(20) , 2004.03(9) , 2004.16(24) , 2004.47(34) , \\
           &       &      &      &  2005.21(13) , 2005.30(2) , 2005.49(8) , 2006.03(12) , 2006.20(4) , \\
           &       &      &      &  2006.38(8) , 2006.44(10) , 2007.07(9) , 2007.23(11) , 2007.47(3) , \\
           &       &      &      &  2008.05(9) , 2008.37(1) , 2008.47(5) \\
  KT Per  &   49  &  82  & 104  & 2003.77(1) , 2004.03(19) , 2004.87(14) , 2005.08(5) , 2005.70(24) , \\
           &       &      &      &  2005.87(12) , 2006.03(14) , 2006.66(5) , 2007.65(2) , 2007.73(8) \\
  VV Pup  &   78  & 131  & 153  & 2003.08(12) , 2004.03(18) , 2004.16(16) , 2004.87(9) , 2005.08(6) , \\
           &       &      &      &  2005.20(10) , 2005.30(8) , 2005.88(12) , 2006.03(14) , 2006.20(16) , \\
           &       &      &      &  2006.83(12) , 2007.07(8) , 2007.23(1) , 2008.05(5) , 2008.15(6) \\
  MR Ser  &   47  &  73  & 107  & 2003.46(1) , 2004.47(16) , 2005.21(10) , 2005.30(9) , 2005.48(10) , \\
           &       &      &      &  2005.71(6) , 2006.20(10) , 2006.37(8) , 2006.44(10) , 2007.23(10) , \\
           &       &      &      &  2007.34(4) , 2007.47(8) , 2008.37(5) \\
  SW UMa  &   19  &  76  & 120  & 1999.05(27) , 2000.03(18) , 2000.26(9) , 2004.03(9) , 2004.16(6) , \\
           &       &      &      &  2004.87(5) , 2005.20(9) , 2005.88(8) , 2006.04(8) , 2006.20(6) , \\
           &       &      &      &  2006.84(8) , 2007.23(7) \\
  AR UMa  &   19  &  44  & 106  & 2003.08(2) , 2005.08(22) , 2005.21(7) , 2005.30(7) , 2005.89(9) , \\
           &       &      &      &  2006.03(11) , 2006.20(10) , 2006.37(8) , 2007.07(4) , 2007.23(10) , \\
           &       &      &      &  2008.04(5) , 2008.14(7) , 2008.37(4) \\
  BZ UMa  &   19  &  50  &  87  & 2000.26(1) , 2002.81(2) , 2004.03(15) , 2005.88(12) , 2006.03(18) , \\
           &       &      &      &  2006.20(10) , 2006.83(3) , 2007.07(8) , 2007.23(10) , 2008.04(8) \\
  QS Vir  &   32  &  72  & 131  & 2003.08(9) , 2003.46(6) , 2004.03(2) , 2004.16(18) , 2004.47(26) , \\
           &       &      &      &  2005.21(14) , 2005.30(10) , 2005.48(6) , 2006.05(1) , 2006.20(14) , \\
           &       &      &      &  2006.37(5) , 2006.44(10) , 2007.23(5) , 2007.35(5) \\

\enddata
\tablecomments{Overview of the data included in the parallax solutions.
$N_{\rm ref}$ is the number
of reference stars used to define the plate solution, $N_{\rm meas}$ is the total
number of stars measured, and  $N_{\rm pix}$ is the number of images used.  The epochs
represent different observing runs, and the numbers in parentheses are the number of
images included from each run.}
\end{deluxetable}

\begin{deluxetable}{rrrrrrrrrr}
\tabletypesize{\scriptsize}
\tablewidth{0pt}
\tablecolumns{5}
\tablecaption{Positions, Magnitudes, Parallaxes, and Proper Motions}
\tablehead{
\colhead{$\alpha$} &
\colhead{$\delta$} & 
\colhead{Weight} &
\colhead{$\sigma$} & 
\colhead{$V$} &
\colhead{$V - I$} & 
\colhead{$\pi_{\rm rel}$} &
\colhead{$\mu_X$} & 
\colhead{$\mu_Y$} &
\colhead{$\sigma_{\mu}$} \\
\colhead{[h:m:s]} & 
\colhead{[$^\circ$:$'$:$''$]} & 
\colhead{} & 
\colhead{[mas]} & 
\colhead{} & 
\colhead{} & 
\colhead{[mas]} & 
\colhead{[mas yr$^{-1}$]} & 
\colhead{[mas yr$^{-1}$]} & 
\colhead{[mas yr$^{-1}$]} \\ 
}
\startdata
DW Cnc: \\
  7:58:51.44 & +16:13:20.6 &  0 &  12 &  18.34 & 0.77  & $-2.3 \pm  1.8$ & $ -3.2 $& $  3.0 $&  1.0 \\ 
  7:58:51.42 & +16:19:32.2 &  1 &   9 &  19.05 & 1.65  & $-0.2 \pm  1.3$ & $ 11.7 $& $ -2.2 $&  0.7  \\
  7:58:52.07 & +16:17:23.0 &  1 &  12 &  18.78 & 0.61  & $ 0.2 \pm  1.8$ & $  1.7 $& $  1.5 $&  1.0  \\
$*$7:58:53.04 & +16:16:45.2 &  0 &   5 &  15.00 & 0.52  & $ 4.0 \pm  0.8$ & $-25.7 $& $  4.8 $&  0.4  \\
  7:58:53.47 & +16:15:18.6 &  1 &   6 &  17.60 & 0.82  & $-0.2 \pm  1.0$ & $ -1.4 $& $ -2.1 $&  0.5  \\
  7:58:54.04 & +16:16:40.4 &  1 &  12 &  18.18 & 0.25  & $ 0.2 \pm  1.8$ & $ -0.2 $& $  1.7 $&  1.0  \\
\enddata
\tablecomments{Parameters for all measured stars in all the fields.
The program star in each field is marked with an asterisk.  
The celestial coordinates are from mean CCD images and are referred
to the USNO A2.0 or UCAC2, which are aligned with the ICRS; the epochs
of the images used are typically around 2005.  Coordinates should
be accurate to $\sim 0''.3$ external and somewhat better than this
internally.   A 1 or 0 in the next
column indicates whether a star was used as a reference star.  The
next column gives the scatter around the best astrometric fit (see
text); very large scatter indicates some kind of problem with the 
centering of that particular star (caused, for example, by 
saturation on some images).  The $V$ and $V-I$ colors come next,
with typical external uncertainties of 0.05 mag.
Next come the fitted parallaxes, proper motions in $X$ and $Y$, 
and the uncertainty in the proper motion (per coordinate). The 
full table available in the electronic version of this paper; a
small sample is shown here as a guide to its form and content.}
\end{deluxetable}
 
\begin{deluxetable}{lrrrr}
% \tabletypesize{\scriptsize}
\tabletypesize{\small}
\tablewidth{0pt}
\tablecolumns{5}
\tablecaption{Parallaxes, Proper Motions, and Distances}
\tablehead{
\colhead{Star} & 
\colhead{$\pi_{\rm rel}$} & 
\colhead{$\pi_{\rm abs}$} & 
\colhead{$[\mu_\alpha, \mu_\delta]_{\rm rel}$} &
\colhead{$1/\pi_{\rm abs}$} \\ 
\colhead{} &
\colhead{$d_{\rm LK}$} &
% \colhead{$d(\pi)$} &
\colhead{$d(\pi,\mu)$} &
\colhead{$(m - M)$ prior} & 
\colhead{$d(\pi, \mu, m-M)$} \\
}
%\colhead{} &
%\colhead{(mas)} &
%\colhead{(mas)} &
%\colhead{(mas yr$^{-1}$)} &
%\colhead{(pc)} &
%\colhead{(pc)} &
%\colhead{(pc)} &
%\colhead{(mag)} &
%\colhead{(pc)} \\
%}
\startdata
  DW Cnc & $4.0 \pm 0.8 [1.0]$  & 4.8(1.0) & $-25.7,+4.8 (0.4)$ & $208(+55,-36)$ \\
 & 268  & $257(+79,-52)$ & \nodata  & $257(+79, -52)$ \\
  HT Cas & $7.9 \pm 0.8 [1.0]$  & 9.0(1.1) & $+27.3,-13.9 (0.6)$ & $111(+16,-12)$ \\
 & 118 & $123(+20,-15)$ & 6.1; 1.0  & $131(+22, -17)$ \\
  MQ Dra & $5.7 \pm 0.7 [1.0]$  & 6.7(1.0) & $-30.8,6.2 (0.6) $ &  $149(+26,-19)$ \\
 & 165 & $167(+35,-25)$ & 5.9; 0.7  & $162(+27, -21)$ \\
  IR Gem & $2.1 \pm 0.6 [1.1]$  & 3.0(1.1) & $+53.9,-22.3(0.3) $ &  $333(+193,-89)$ \\
 & \nodata & $413(+138,-112)$ & 11.7; 4.6; 1.0  & $358(+104, -86)$ \\
V396 Hya & $10.3 \pm 1.2 [1.2]$ & 11.1(1.2) & $-276.7,-52.4(0.7)$ & $90(+11,-9)$ \\
 & 94 &  $92 (+13,-10)$ & 17.6; 13.5; 1. & $90(+12,-10)$ \\ 
KT Per & $5.8 \pm 0.9 [1.2]$  & 6.9(1.2) & $58.1,+0.2 (0.8)$ & $145(+30,-22)$ \\
 & 172 & $162 (+36, -26)$ & 6.9; 0.7\tablenotemark{a} & $180 (+36,-28)$  \\
VV Pup & $8.8 \pm 1.4 [1.2]$ & 9.3; 1.2 & $+19.0, -70.4 (0.8)$ & $108(+16,-12)$ \\
 & 115 & $114 (+18, -14)$ & 5.9; 0.6 & $124 (+17, -14)$ \\
MR Ser & $8.4 \pm 0.8 [1.0]$ & 9.2(1.0) & $-35.7, +65.6 (0.6)$ & $109(+13,-11)$ \\
 & 114 & $113 (+14, -12)$ & 6.0; 0.5 & $126 (+14, -12)$ \\
SW UMa & $5.6 \pm 0.8 [1.2]$ & 6.6(1.2) & $-31.6, +8.9 (0.2)$ & $152(+34,-23)$ \\
 & 193 & $180 (+49, -33)$ & 6.0; 0.4 & $164 (+22,-19)$ \\
AR UMa & $11.2 \pm 1.2 [1.2]$ & 12.2(1.2) & $-68.5, +3.4 (0.8)$ & $82(+9,-7)$ \\
 &  85 & $86 (+10, -8)$ & 4.7; 0.6 & $86 (+9, -8)$ \\
BZ UMa & $3.8 \pm 0.9$ & 4.9(1.1) & $21.9, -5.8 (0.5)$ & $204 (+59, -37)$ \\
 & \nodata & $260 (+93, -58)$ & 5.2; 1.4 & $228 (+63, -43)$ \\
QS Vir & $20.6 \pm 1.7 [1.7]$ & 21.2(1.7) & $50.1, 19.9 (1.1)$ & $47 (+4, -4)$ \\
 & 48 & $49 (+5, -4)$ & 3.26; 1.0 & $49 (+4, -4)$ \\
% checked up to here - 2008 Jul 15
\enddata
\clearpage
\tablecomments{Summary of the parallaxes, proper motions, prior information, 
and distance estimates.  
Two lines are given for each star, as follows.  {\it Line 1.} 
Column 1: Name of star.  Column 2: relative parallax and its uncertainty; 
the first uncertainty given is from the formal error of the fit, 
and the estimated external error is given in square brackets.  Column 3: 
Adopted absolute parallax
and its error (in parentheses).  Column 4:  Relative proper motion 
in $X$ (i.e., eastward) and $Y$ (northward), and the formal error per 
coordinate in parentheses.  Column 5: Distance implied by the absolute
parallax, and its uncertainty, with no further corrections applied.
{\it Line 2.}  Column 1: Left blank.  Column 2: 50th percentile 
of the distance probability distribution after the Lutz-Kelker correction
is applied.  Column 3: 50th percentile distance estimate using the 
parallax and proper motion information, and a model for the intrinsic
CV velocity distribution.  The uncertainties given are the 16th and
84th percentile points, equivalent to $\pm 1 \sigma$.  Column 4:
Prior distance estimate.  Where two numbers are given, they are the
distance modulus $m - M$ and its assumed uncertainty; where there are
three numbers, they are an apparent magnitude $m$, an absolute
magnitude $M$, and an uncertainty in $m - M$.  Column 5: 
The 50th percentile Bayesian distance estimate, including all the 
prior information.  Uncertainties are again 16th and 84th percentile 
points. 
}
\tablenotetext{a}{The photometric distance constraint for KT Per is from the
spectroscopic parallax of the common-proper-motion companion; see text for
details.}
\label{tab:results}
\end{deluxetable}

\end{document}